# Planar fiber-chip-coupling using angle-polished polarization maintaining fibers

M. Schneider[a,1], L. A. Garcia Herrera[a], B. Burger[a], L. Eisenblätter[a], T. Kühner[a]

[a] *Karlsruhe Institute of Technology, Institute for Data Processing and Electronics, 76131 Karlsruhe, Germany*
 *E-mail*: marc.schneider@kit.edu

ABSTRACT: We report on our latest developments of a planar fiber-chip-coupling scheme, using angle polished, polarization maintaining (PM) fibers. Most integrated photonic chip components are polarization sensitive and a suitable way to launch several wavelength channels with the same polarization to the chip is the use of PM fibers. Those impose several challenges at processing and handling to achieve a stable, permanent, and low-loss coupling. We present the processing of the fibers in detail and experimental results for our planar and compact fiber-chip-coupling technique.

KEYWORDS: Glass fibers, polarization maintaining fibers, fiber processing, fiber-chip-coupling, angle-polishing, grating couplers.

## Contents

## 1. Introduction

High performance optical links using wavelength division multiplexing (WDM) are the future in detector instrumentation to increase data transmission bandwidth and reduce fiber count. A key issue of such a system is a compact and efficient fiber-chip-coupling, connecting optical glass fibers to the photonic chip. As most components of integrated photonic chips are polarization sensitive, one can use lossy on-chip polarization-insensitive couplers and polarization controllers for each wavelength or use polarization maintaining (PM) fibers to feed the required polarization directly from the lasers without the need of further manipulation.

Often used on-chip components for light-coupling are grating couplers, which offer the advantage to couple light in to and out of the chip-integrated waveguides almost anywhere on the chip and not just at the edges. For coupling, grating couplers are irradiated with light under a certain angle to the chip's normal. Commonly used angles are around 10°, but they depend on the specific design of the grating coupler and the used wavelength. Launching light with a cleaved fiber leads to a rather high buildup, as the minimum bending radius of standard glass fibers is in the range of several tens of millimeters [1]. For smaller, more planar buildups, other techniques are required.

Our fiber-chip-coupling uses optical single mode glass fibers, whose tip is polished to a certain angle, so that light is reflected radially out of the fiber by total internal reflection at a defined angle (Figure 1). The fiber is positioned parallel to the chip surface with the tip above an on-chip grating coupler, so that the radially emitted light hits the grating coupler, which diffracts the light into an on-chip waveguide for further routing [2, 3].

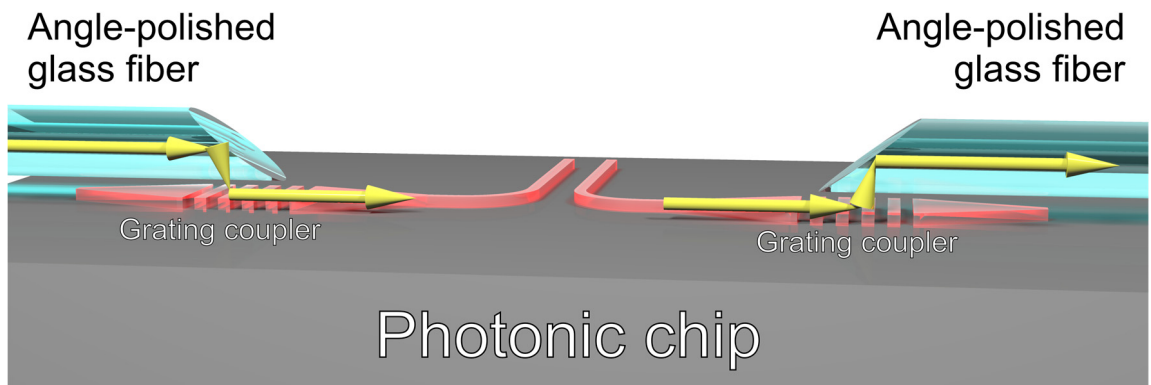

**Figure 1.** Scheme of fiber-chip-coupling using grating couplers on chip and angle-polished glass fibers.

PM fibers, however, are not axially symmetric due to the strain rods close to the fiber core introducing birefringence (Figure 2). Therefore fiber-chip-coupling becomes more challenging, as the angle of in-axis rotation has to be very well defined.

We will present the setup to initially adjust the PM fibers for polishing, the optimized polishing process, a recipe to reproduce the results, as well as measurement results of fiber-chip-coupling experiments.

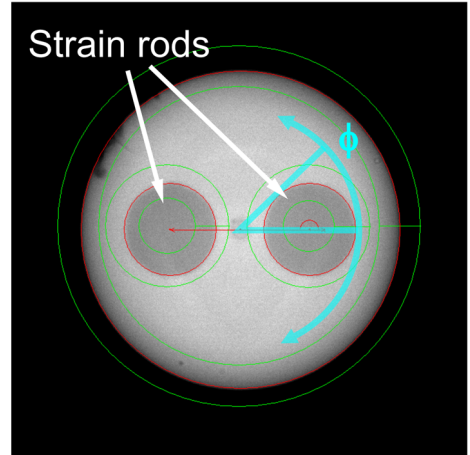

**Figure 2.** Microscope image of tip of cleaved polarization maintaining fiber.

## 2. Fiber Processing

Based on the work published in [3], we refined and enhanced the fiber processing for polarization maintaining fibers. For a low-loss fiber chip coupling, the position and angular errors have to be kept as small as possible. In our process the angle of axial rotation $\phi$ is adjusted for the polishing process with an error of less than 0.3°. According to our simulations, this value should keep the additional coupling losses as low as 0.02 dB, which is negligible compared to other sources like deviation from the ideal polishing angle and wavelength dependence of the grating couplers. To achieve this, a special setup was developed (Figure 3), which allows a precise fiber rotation before gluing the fibers to a fiber holder for polishing. The fiber holder consists of a grooved microscope slide. A cleaved fiber with clean front facet is placed into one of the grooves and, further back, in a fiber rotator from Piezosystem Jena. For alignment, the image of the front facet of a cleaved PM fiber with its core, cladding, and, most important, strain rods is captured by a camera with a microscope objective, which is placed on a XYZ-linear stage. The camera picture is analyzed by a machine vision system, that extracts the contours of the fiber and the strain rods, calculates the rotation angle and indicates, if the angle is in an acceptable range. To display the glass strain rods inside the cladding glass, special lighting with gracing incidence from behind, stray light shading, and digital contrast enhancement is required.

If the rotation angle is correct, a small drop of UV-curing glue (Delo Photobond GB368) is applied to fix the fiber. Additional glue is applied afterwards to improve fixation. The process is repeated for several more fibers until all grooves are occupied.

The following polishing process is essential for a low-loss coupling, as the angle of the polished surface to the fiber axis and the surface quality are equally important. To maintain the polishing angle, a special polishing fixture with a parallelogram guidance was developed and printed (Figure 4) using a Makerbot Replicator Dual 3D-printer. The polishing angle is defined by the angle of the main, wedge shaped part of the chuck.

Several silicon carbide lapping films and diamond suspensions with decreasing grain sizes were used for polishing. Table 1 lists the used materials with grain sizes, polishing durations, and rotation speeds of the polishing machine used for each step. Between each step, the fiber chuck and the machine were thoroughly cleaned to avoid contamination with coarser grains. Figure 5 shows the polishing chuck on the disk of our polishing machine. The chuck was placed on the rim and the moving part was pushed onto the rotating disk by hand. Here it is important that the applied pressure is roughly similar for each batch of fibers. As there was, due to the applied pressure, a slight discrepancy of 1.3° between the desired angle set by the polishing chuck and the angle of the polished surface obtained, the angle of the chuck was optimized in several iterations to obtain fibers with a coupling angle of 12°, which is the optimal angle for a center wavelength of 1550 nm of the grating couplers on our photonic chips.

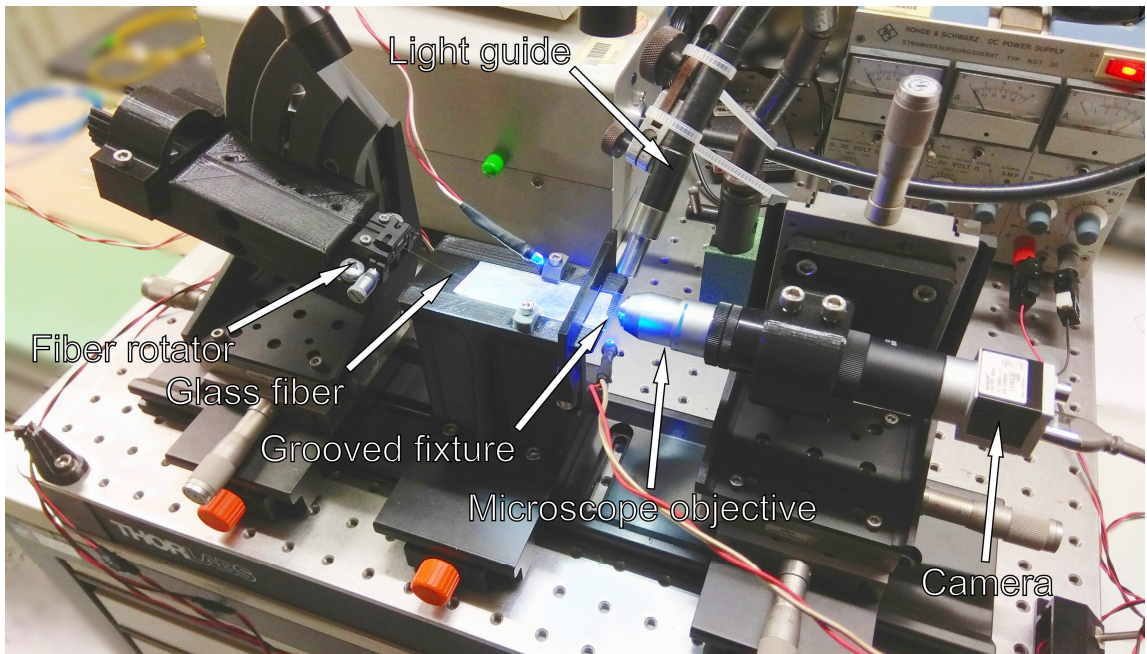

**Figure 3.** Setup for axial alignment of PM fibers in grooved holder for polishing.

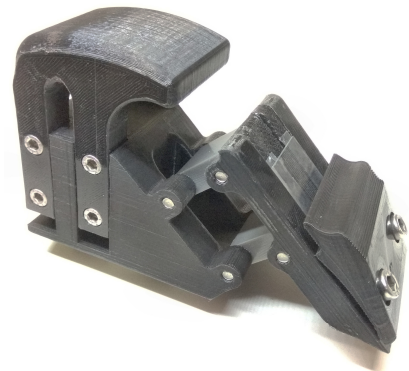

After polishing, the fibers were removed from their holder by immersion in acetone for 30 minutes. The used glue swells in acetone and loses most of its adhesion, but removal has to be careful anyway, not to break the fibers.

**Figure 4.** Fiber polishing chuck.

Residues were wiped away with clean isopropanol. The final step to achieve a really smooth surface was melting the glass surface. This was done by installing the angle polished fiber in a fusion splicer and starting a standard program for fusing single mode fibers. Most fusion splicers clean the surface of the respective fibers first by a short electric arc, burning residues and melting the front facet to achieve good splicing results. For the angle polished fiber, the splicing program was cancelled after this first step and the fiber removed.

**Table 1.** Polishing sequence and parameters.

| Material | Grain size | Duration | Rotation speed |
|---|---|---|---|
| Lapping film | 15 µm | 1 min. | 200 rpm |
| Lapping film | 8 µm | 5 min. | 200 rpm |
| Diamond emulsion | 3 µm | 8 min. | 150 rpm |
| Diamond emulsion | 1 µm | 8 min. | 150 rpm |
| Diamond emulsion | 0.25 µm | 8 min. | 150 rpm |

The result of this procedure is shown for one fiber in Figure 6. The front facet has a smooth mirror finish without scratches or debris.

## 3. Results

After processing, the fibers were characterized by comparative coupling experiments. For this, reference measurements were made with straight waveguides, terminated with grating couplers on a silicon photonic chip. The used fibers were cleaved single mode glass fibers, mounted at the optimal coupling angle to achieve a minimum coupling loss of 9.8 dB at a center wavelength of 1550 nm (Figure 7). The waveguide loss was found to be 0.6 dB and the coupling loss for a single grating coupler to be 4.6 dB. As the reference spectrum in Figure 7 belongs to a pair of couplers with a superposition of both transmission spectra, the 3dB-bandwidth of a single coupler can be found 6 dB below the maximum, indicated by a horizontal dotted line, with 76.2 nm. With this reference, one fiber was replaced by an angle polished fiber to be characterized. The angle polished fiber was placed into a fiber rotator similar to the one used in the fiber preparation setup. Special care was spent to align the fiber axis parallel to the chip surface. In several iterations, the optimal rotation angle was determined and, for standard fibers, the optical polarization optimized by a polarization controller. With the optimum angle, a transmission spectrum was measured. The results were corrected by the reference spectrum to determine the additional coupling loss for the angle polished glass fibers.

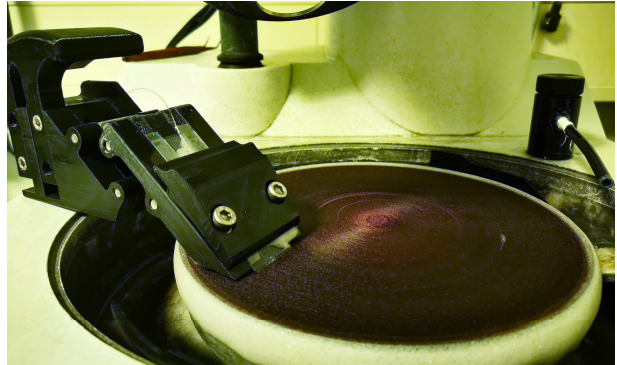
**Figure 5.** Fiber polishing on polishing machine with fiber chuck.

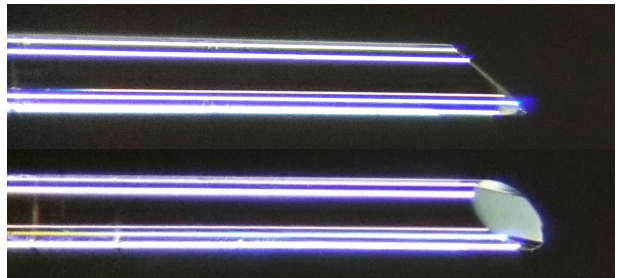
**Figure 6.** Angle polished PM fiber.

Figure 8a) shows the additional loss for four angle-polished standard glass fibers over the wavelength. The dots show the processed measurement data, the lines a polynomial fit to guide the eye. As can be seen, the measurements scatter up to 4 dB, which is a result of some back reflections and interference in the waveguide as well as between grating couplers and fibers. The line fits reveal a much clearer picture and show that there is in general no additional loss with respect to coupling with cleaved fibers.

Similar measurements for PM fibers are shown in Figure 8b). Again, the points are the processed measurement data, the lines polynomial fits. Here the spread is much larger and the wave-like shape of the fit curves, most prominently for fiber 1, shows a transmission max-

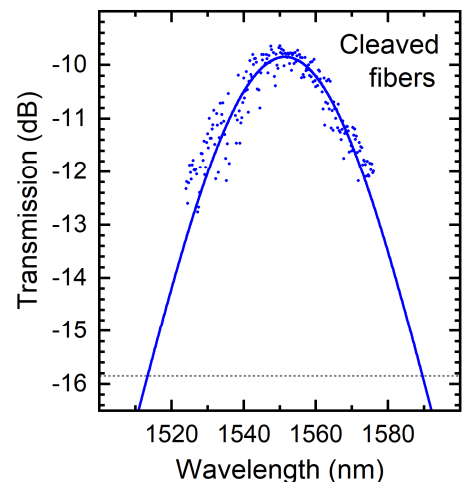
**Figure 7.** Reference transmission spectrum of two cleaved fibers above grating couplers on photonic chip.

imum other than 1550 nm (for fiber 1 the transmission maximum was measured to be 1545 nm). These fibers show a larger deviation in coupling angles as well as a much higher loss. On average the additional coupling loss for PM fibers is 4.2 dB with a standard deviation of 1.2 dB, while the absolute spread is +3.4 dB and -2.4 dB. The higher loss was not unexpected, as we observed a similar penalty over standard fibers using cleaved PM fibers for coupling. A possible reason for the higher loss might be a significant mode mismatch, but that needs further investigation.

## 4. Conclusions

We presented a method to adjust PM fibers for angle-polishing and a recipe to polish them with high quality for a planar fiber-chip-coupling using on-chip grating couplers. The polishing process contains several mechanical polishing steps, followed by smoothing the polished surface by melting with an electric arc. Consecutive measurements of such processed fibers show that they exhibit no additional coupling loss, but coupling with PM fibers still suffers from a 4.2 dB higher loss compared to standard fibers. To reduce it, further investigations and optimizations are required.

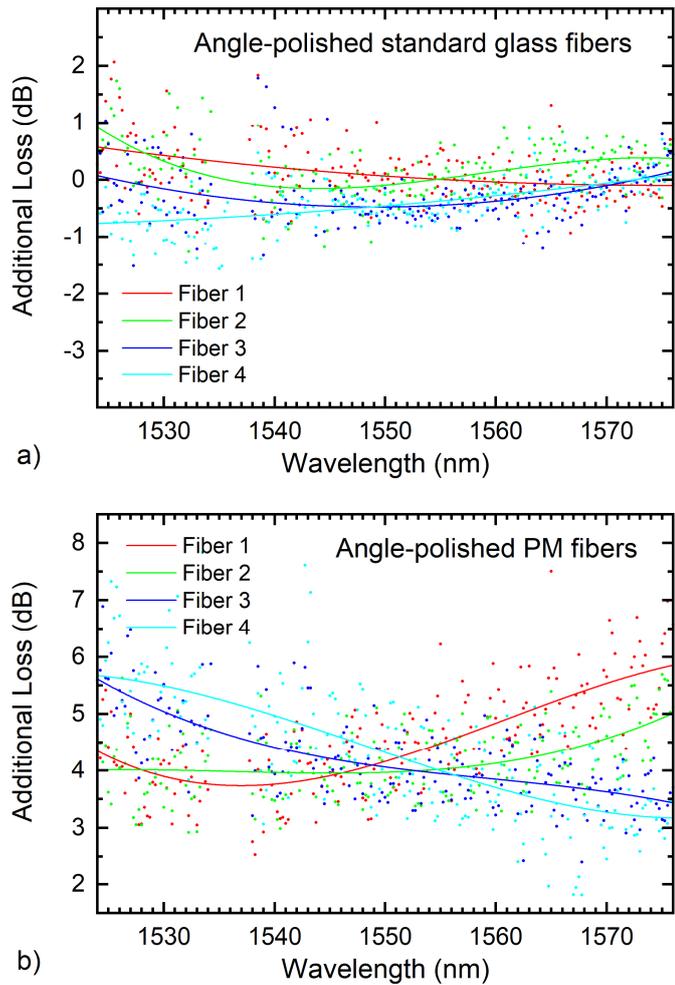

**Figure 8.** Measurement results of additional coupling loss for a) angle-polished standard fibers and b) angle-polished PM fibers.

## References

[1] Datasheet of Corning SMF-28e+ single mode glass fiber, viewed at 2022-10-17, https://www.corning.com/media/worldwide/coc/documents/Fiber/PI-1463-AEN.pdf

[2] D. Karnick et al., *Optical Links for Detector Instrumentation: On-Detector Multi-Wavelength Silicon Photonic Transmitters*, JINST, DOI 10.1088/1748-0221/12/03/C03078.

[3] D. Karnick et al., *Efficient, easy-to-use, planar fiber-to-chip coupling process with angle-polished fibers*, 67th ECTC, DOI 10.1109/ECTC.2017.245.